\documentclass[11pt]{article}
\begin{document}

\begin{center}
\large\textbf{Remarks on remnants by fermions' tunnelling from black strings}
\end{center}

\begin{center}
Deyou Chen \footnote{ E-mail: \underline
{dyouchen@gmail.com} }  and Zhonghua Li \footnote{ E-mail: \underline
{sclzh888@163.com} }
\end{center}

\begin{center}
Institute of Theoretical Physics, China West Normal University, \\Nanchong 637009, China
\end{center}

\textbf{Abstract:} Hawking's calculation is unable to predict the final stage of the black hole evaporation. When effects of quantum gravity are taken into account, there is a minimal observable length. In this paper, we investigate fermions' tunnelling from the charged and rotating black strings. With the influence of the generalized uncertainty principle, the Hawking temperatures are not only determined by the rings, but also affected by the quantum numbers of the emitted fermions. Quantum gravity corrections slow down the increases of the temperatures, which naturally leads to remnants left in the evaporation.

\section{Introduction}

Hawking radiation is a quantum tunnelling phenomenon of particles across black holes' horizons. To describe this phenomenon, the semi-classical tunnelling method, which relies on calculating the imaginary part of a emission particle's action, was put forward \cite{KW}. Adopting the WKB approximation, one can get the relationship between the tunnelling rate and the action of the classically forbidden trajectory of the particle. Here we adopt the canonically invariant expression \cite{APGS,APGS1,APGS2}

\begin{eqnarray}
\Gamma \propto exp\left[-Im\oint p dr\right].
\label{eq1.1}
\end{eqnarray}

\noindent This canonically invariant relation was first derived in \cite{ETA1,ETA2,ETA3}.

The null geodesic method and the Hamilton-Jacobi method are usual methods employed to derive the imaginary part \cite{PW,SP,SPS,ANVZ,KM1}. In the null geodesic method \cite{PW}, we should first perform the Painleve coordinate transformation on a metric. Then use canonical momenta and Hamilton canonical equations to get the imaginary part. When the variable background spacetime is taken into account, the corrected Hawking temperature is higher than the standard one. Therefore, the variable background spacetime implies the accelerated evaporation. The equation of motion of a massive particle is different from that of the massless one. The former obeys de Broglie wave function relation. Therefore, the phase velocity of the particle was adopted to research the tunnelling radiation of massive particles in the subsequent investigations \cite{ZZ,JWC}. The embryonic form of the the Hamilton-Jacobi method \cite{ANVZ} was first found in \cite{SP,SPS}. In this method, the action satisfies the Hamilton-Jacobi equation. Taking into account the property of the spacetime, one carries out separation of variables on the action. Then inserting the separated variables into the Hamilton-Jacobi equation and solving it, one gets the imaginary part. Extending this work to the tunnelling radiation of fermions, the standard Hawking temperatures of the spherically symmetric and charged black holes were recovered  \cite{KM2}. Other work about fermions' tunnelling radiation is referred to \cite{KM3,AS,AS1,LR,CJZ,QQJ,HCNVZ,HCNVZ1,BRM}. The standard temperatures were also recovered by anomaly cancellations \cite{RW,RW1,BM01}.

Various theories of quantum gravity predict the existence of a minimal observable length \cite{PKTC,PKTC1,PKTC2,PKTC3,PKTC4}. This length can be implemented in the model of the generalized uncertainty principle (GUP)

\begin{eqnarray}
\Delta x \Delta p \geq \frac{\hbar}{2}\left[1+ \beta \Delta p^2\right],
\label{eq1.1}
\end{eqnarray}

\noindent where $\beta = \beta_0 \frac{l^2_p}{\hbar^2}$, $\beta_0 $ is a dimensionless parameter and $l_p$ is the Planck length. The derivation of the GUP is relied on the modified fundamental commutation relations. Kempf et. al. first modified commutation relations \cite{KMM} and got $\left[x_i,p_j\right]= i \hbar \delta_{ij}\left[1+ \beta p^2\right]$, where $x_i$ and $p_i$ are operators of position and momentum defined by

\begin{eqnarray}
x_i &=& x_{0i}, \nonumber\\
p_i &=& p_{0i} (1 + \beta p^2),
\label{eq1.2}
\end{eqnarray}

\noindent and $x_{0i}$ and $p_{0i}$ satisfy the canonical commutation relations $\left[x_{0i},p_{0j}\right]= i \hbar \delta_{ij}$.

This modification plays an important role in quantum gravity. With considerations of modifications, the cosmological constant problem was discussed and the finiteness of the constant was derived in \cite{CMOT}. Using a new form of GUP, the Unruh effect has been analyzed in \cite{BM06}. The quantum dynamics of the Friedmann-Robertson-Walker universe was gotten in \cite{BM1}. The related predictions on post inflation preheating in the cosmology were derived in \cite{CDAV}. Using the modifications, the thermodynamics of the black holes were researched in \cite{KSY,XW,BJM} and the tunnelling radiation of scalar particles was investigated in \cite{NS}. In recent work \cite{CWY}, taking into account effects of quantum gravity, the authors modified the Dirac equation in curved spacetime and investigated fermions' tunnelling from the Schwarzschild black hole. They derived that the quantum correction slows down the increase of the Hawking temperature, which leads to the remnant.

In this paper, we extend this work to anti de Sitter spacetimes and investigate the tunnelling radiation of fermions from black strings, where effects of quantum gravity are taken into account. Black strings are cylindrically symmetric solutions of the Einstein-Maxwell equations with a negative cosmological constant. The solutions are asymptotically anti de Sitter in transverse direction and along the axis. There are three Killing vectors, $\partial_t$, $\partial_{\theta}$, $\partial_z$ as the minimal symmetry. The AdS/CFT correspondence is an important topic in modern physics. Researches of anti de Sitter spacetimes are helpful to understand this correspondence. To incorporate effects of quantum gravity, we first modify the Dirac equation in curved spacetime by operators of position and momentum defined in \cite{KMM}. Then adopt the Hamilton-Jacobi method to get the imaginary parts of the actions. The corrected Hawking temperatures are not only determined by the mass, charge and angular momentum of the strings, but also affected by the quantum numbers (charge, angular momentum, mass and energy) of the emitted fermions. Quantum gravity corrections slow down the increases of the Hawking temperatures. It is natural to lead to the remnants left in the evaporation.

The rest is organized as follows. In the next section, using operators of position and momentum defined in \cite{KMM}, we modify the Dirac equation in curved spacetime. In Sect. 3, we investigate the tunnelling radiation of charged fermions from the charged black string. The remnant is observed in the evaporation. In Sect. 4, the radiation of uncharged fermions in the rotating black string is discussed. Sect. 5 is devoted to our conclusion.

\section{Generalized Dirac equation}

Here we adopt the modified fundamental commutation relation put forward in \cite{KMM} to modify the Dirac equation in curved spacetime. Using Eq. (\ref{eq1.2}), the square of momentum operators is gotten as

\begin{eqnarray}
p^2 &=& p_i p^i = -\hbar^2 \left[ {1 - \beta \hbar^2 \left( {\partial _j \partial ^j} \right)}
\right]\partial _i \cdot \left[ {1 - \beta \hbar^2 \left( {\partial ^j\partial _j } \right)}
\right]\partial ^i\nonumber \\
&\simeq & - \hbar ^2\left[ {\partial _i \partial ^i - 2\beta \hbar ^2
\left( {\partial ^j\partial _j } \right)\left( {\partial
^i\partial _i } \right)} \right].
\label{eq2.1}
\end{eqnarray}

\noindent The higher order terms of $\beta $ are neglected in the last step. In the theory of quantum gravity, the generalized frequency is found as \cite{WG}

\begin{eqnarray}
\tilde \omega = E( 1 - \beta E^2),
\label{eq2.2}
\end{eqnarray}

\noindent where $E$ is the energy operator and defined as $ E = i \hbar \partial _t $. From the energy mass shell condition $ p^2 + m^2 = E^2 $, the generalized expression of the energy was derived \cite{NS,WG,NK1,HBHRSS}. It is

\begin{eqnarray}
\tilde E = E[ 1 - \beta (p^2 + m^2)].
\label{eq2.3}
\end{eqnarray}

\noindent The generalized Dirac equation without considerations of electromagnetic effects in the flat spacetime has been derived in \cite{NK1} by the consequence of the GUP. In curved spacetime, the Dirac equation with an electromagnetic field takes on the form

\begin{eqnarray}
i\gamma^{\mu}\left(\partial_{\mu}+\Omega_{\mu}+\frac{i}{\hbar}eA_{\mu}\right)\Psi+\frac{m}{\hbar}\Psi=0,
\label{eq2.4}
\end{eqnarray}

\noindent where $\Omega _\mu \equiv\frac{i}{2}\omega _\mu\, ^{a b} \Sigma_{ab}$, $\omega _\mu\, ^{ab}$ is the spin connection defined  by the tetrad $e^\lambda\,_b$ and ordinary connection

\begin{eqnarray}
\omega_\mu\,^a\,_b=e_\nu\,^a e^\lambda\,_b \Gamma^\nu_{\mu\lambda} -e^\lambda\,_b\partial_\mu e_\lambda\,^a.
\label{eq2.5}
\end{eqnarray}

\noindent The Latin indices live in the flat metric $\eta_{ab}$ while Greek indices are raised and lowered by the curved metric $g_{\mu\nu}$. The tetrad can be constructed from

\begin{eqnarray}
g_{\mu\nu}= e_\mu\,^a e_\nu\,^b \eta_{ab},\hspace{5mm} \eta_{ab}= g_{\mu\nu} e^\mu\,_a e^\nu\,_b, \hspace{5mm} e^\mu\,_a e_\nu\,^a=\delta^\mu_\nu, \hspace{5mm} e^\mu\,_a e_\mu\,^b = \delta_a^b.
\label{eq2.6}
\end{eqnarray}

\noindent In equation (\ref{eq2.4}), $\Sigma_{ab}$ is the Lorentz spinor generators defined by

\begin{equation}
\Sigma_{ab}= \frac{i}{4}\left[ {\gamma ^a ,\gamma^b} \right], \hspace{5mm} \{\gamma ^a ,\gamma^b\}= 2\eta^{ab}.
\label{eq2.7}
\end{equation}

\noindent Then one can construct $\gamma^\mu$'s in the curved spacetime as

\begin{equation}
\gamma^\mu = e^\mu\,_a \gamma^a, \hspace{7mm} \left\{ {\gamma ^\mu,\gamma ^\nu } \right\} = 2g^{\mu \nu }.
\label{eq2.8}
\end{equation}

\noindent To get the generalized Dirac equation in the curved spacetime, we rewrite Eq. (\ref{eq2.4}) as

\begin{eqnarray}
-i\gamma^{0}\partial_{0}\Psi=\left(i\gamma^{i}\partial_{i}+i\gamma^{\mu}\Omega_{\mu}+i\gamma^{\mu}\frac{i}{\hbar}eA_{\mu}+\frac{m}{\hbar}\right)\Psi.
\label{eq2.9}
\end{eqnarray}

\noindent Using Eqs. (\ref{eq2.1}), (\ref{eq2.3}) and (\ref{eq2.9}) and neglecting the higher order terms of $\beta$, we get \cite{NS,WG,NK1,HBHRSS}

\begin{eqnarray}
-i\gamma^{0}\partial_{0}\Psi=\left(i\gamma^{i}\partial_{i}+i\gamma^{\mu}\Omega_{\mu}+i\gamma^{\mu}
\frac{i}{\hbar}eA_{\mu}+\frac{m}{\hbar}\right)\left(1+\beta\hbar^{2}\partial_{j}\partial^{j}-\beta m^{2}\right)\Psi,
\label{eq2.10}
\end{eqnarray}

\noindent which is rewritten as

\begin{eqnarray}
\left[i\gamma^{0}\partial_{0}+i\gamma^{i}\partial_{i}\left(1-\beta m^{2}\right)+i\gamma^{i}\beta\hbar^{2}\left(\partial_{j}\partial^{j}\right)\partial_{i}+\frac{m}{\hbar}
\left(1+\beta\hbar^{2}\partial_{j}\partial^{j}-\beta m^{2}\right)\right.\nonumber \\
\left.+i\gamma^{\mu}\frac{i}{\hbar}eA_{\mu}\left(1+\beta\hbar^{2}\partial_{j}\partial^{j}-\beta m^{2}\right)+i\gamma^{\mu}\Omega_{\mu}\left(1+\beta\hbar^{2}\partial_{j}\partial^{j}-\beta m^{2}\right)\right]\Psi= 0.
\label{eq2.11}
\end{eqnarray}

\noindent Thus the generalized Dirac equation is derived. When $A_{\mu} = 0$, it describes a equation without electromagnetic fields. In the following sections, we adopt Eq. (\ref{eq2.11}) to describe fermions tunnelling from the charged and rotating black strings.

\section{Fermions' tunnelling from a charged black string}

The 4-dimensional neutral black string solutions to Einstein-Maxwell equations with a negative cosmological constant were derived in \cite{JPSL}. Subsequently, the general solutions with electric charges were gotten \cite{CZ}. In this section, we investigate charged fermions' tunnelling from a cylindrically symmetric black string. The black string solution is given by \cite{CZ}

\begin{eqnarray}
ds^{2}&=&  -f(r)dt^2+ \frac{1}{g(r)}dr^2 + r^2d{\theta}^2 + \alpha^2r^2dz^2,
\label{eq3.1}
\end{eqnarray}

\noindent with the electromagnetic potential

\begin{eqnarray}
A_{\mu} = \left(A_{t},0,0,0\right) = \left(\frac{2Q}{\alpha r},0,0,0\right),
\label{eq3.2}
\end{eqnarray}

\noindent where $f(r)= g(r)=\alpha^2r^2-\frac{4M}{\alpha r} + \frac{4Q^2}{\alpha ^2 r^2}$, $0\leq \theta \leq 2\pi$,  $\alpha ^2=-\frac{\Lambda}{3}$, $\Lambda$ is the negative cosmology constant. $M$ and $Q$ are the ADM mass and charge per unit length in the $z$ direction, respectively. The above spacetime is asymptotically anti-de Sitter in the transverse directions and string directions. The singularity at $r = 0$ is enclosed by the horizon $r_{+}$ if the condition $Q^2\leq \frac{3}{4}M^{\frac{4}{3}}$ holds. The event horizon $r_+$ is located at

\begin{eqnarray}
r_{+} = \frac{1}{2}\left[\sqrt{2R} + \left( -2R + \frac{8M}{\alpha^3\sqrt{2R}}\right)\right],
\label{eq3.3}
\end{eqnarray}

\noindent where

\begin{eqnarray}
R = \left[ \frac{M^2}{\alpha^6} + \left(\left( \frac{M^2}{\alpha^6}\right)^2- \left( \frac{4Q^2}{3\alpha^4}\right)^3\right)^{\frac{1}{2}} \right]^{\frac{1}{3}}
+ \left[ \frac{M^2}{\alpha^6} - \left(\left( \frac{M^2}{\alpha^6}\right)^2- \left( \frac{4Q^2}{3\alpha^4}\right)^3\right)^{\frac{1}{2}} \right]^{\frac{1}{3}}.
\label{eq3.4}
\end{eqnarray}

\noindent The metric (\ref{eq3.1}) describes a neutral black string solution when $Q = 0$.

For a spin-1/2 fermion, there are two states corresponding to spin up and spin down. Here we only investigate the state with spin up. The investigation of the state with spin down is parallel and the same result can be obtained. To describe the motion of a charge fermion, we suppose that the wave function takes on the form

\begin{eqnarray}
\Psi=\left(\begin{array}{c}
A\\
0\\
B\\
0
\end{array}\right)\exp\left(\frac{i}{\hbar}I\left(t,r,\theta , z \right)\right),
\label{eq3.5}
\end{eqnarray}

\noindent where $A$ and $B$ are functions of $ t, r,  \theta , z$, and $I$ is the action of the fermion with spin up state. To find gamma matrices, we should first construct a tetrad. It is straightforward to construct a tetrad from the metric (\ref{eq3.1}). The tetrad is

\begin{eqnarray}
e_\mu\,^a = \rm{diag}\left(\sqrt f, 1/\sqrt g, r,\alpha r \right).
\label{eq3.6}
\end{eqnarray}

\noindent Then gamma matrices are gotten as

\begin{eqnarray}
\gamma^{t}=\frac{1}{\sqrt{f\left(r\right)}}\left(\begin{array}{cc}
i & 0\\
0 & -i
\end{array}\right), &  & \gamma^{\theta}=\sqrt{g^{\theta\theta}}\left(\begin{array}{cc}
0 & \sigma^{1}\\
\sigma^{1} & 0
\end{array}\right),\nonumber \\
\gamma^{r}=\sqrt{g\left(r\right)}\left(\begin{array}{cc}
0 & \sigma^{3}\\
\sigma^{3} & 0
\end{array}\right), &  & \gamma^{z}=\sqrt{g^{zz}}\left(\begin{array}{cc}
0 & \sigma^{2}\\
\sigma^{2} & 0
\end{array}\right).
\label{eq3.7}
\end{eqnarray}

\noindent In the above equations, $\sqrt{g^{\theta\theta}} = \frac{1}{r}$ and $\sqrt{g^{zz}} = \frac{1}{\alpha r}$. To apply the WKB approximation, we insert the wave function and the gamma matrices into the generalized Dirac equation. Then divide by the exponential term and multiply by $\hbar$. The resulting equation to leading order in $\hbar$ is derived and decoupled into four equations

\begin{eqnarray}
-iA\frac{1}{\sqrt{f}}\partial_{t}I-B\left(1-\beta m^{2}\right)\sqrt{g}\partial_{r}I-Am\beta\left[g^{rr}\left(\partial_{r}I\right)^{2}+g^{\theta\theta}\left(\partial_{\theta}I\right)^{2}+
g^{zz}\left(\partial_{z}I\right)^{2}\right]\nonumber\\
+B\beta\sqrt{g}\partial_{r}I\left[g^{rr}\left(\partial_{r}I\right)^{2}+g^{\theta\theta}\left(\partial_{\theta}I\right)^{2}+
g^{zz}\left(\partial_{z}I\right)^{2}\right]+Am\left(1-\beta m^{2}\right)\nonumber\\
-iA\frac{eA_t}{\sqrt f}\left[1-\beta m^{2} -\left(g^{rr}\left(\partial_{r}I\right)^{2}+g^{\theta\theta}\left(\partial_{\theta}I\right)^{2}+
g^{zz}\left(\partial_{z}I\right)^{2}\right)\right]
= 0,
\label{eq3.8}
\end{eqnarray}

\begin{eqnarray}
iB\frac{1}{\sqrt{f}}\partial_{t}I-A\left(1-\beta m^{2}\right)\sqrt{g}\partial_{r}I-Bm\beta\left[g^{rr}\left(\partial_{r}I\right)^{2}+g^{\theta\theta}\left(\partial_{\theta}I\right)^{2}+
g^{zz}\left(\partial_{z}I\right)^{2}\right]\nonumber\\
+A\beta\sqrt{g}\partial_{r}I\left[g^{rr}\left(\partial_{r}I\right)^{2}+g^{\theta\theta}\left(\partial_{\theta}I\right)^{2}+
g^{zz}\left(\partial_{z}I\right)^{2}\right]+Bm\left(1-\beta m^{2}\right)\nonumber\\
+iB\frac{eA_t}{\sqrt f}\left[1-\beta m^{2} -\left(g^{rr}\left(\partial_{r}I\right)^{2}+g^{\theta\theta}\left(\partial_{\theta}I\right)^{2}+
g^{zz}\left(\partial_{z}I\right)^{2}\right)\right]
= 0,
\label{eq3.9}
\end{eqnarray}

\begin{eqnarray}
A\left\{-\left(1-\beta m^2\right) \sqrt {g^{\theta \theta}}\partial _ {\theta} I
+ \beta \sqrt {g^{\theta \theta}}\partial _ {\theta}I\left[g^{rr}(\partial _r I)^2 + g^{\theta \theta}(\partial _ {\theta}I)^2
+ g^{zz}(\partial _ {z}I)^2 \right]\right.\nonumber\\
\left.-i\left(1-\beta m^2\right) \sqrt {g^{zz}}\partial _ {z}I+ i\beta \sqrt {g^{zz}}\partial _ {z}I\left[g^{rr}
(\partial _r I)^2 + g^{\theta \theta}(\partial _ {\theta}I)^2 + g^{zz}(\partial _ {z}I)^2\right]\right\} = 0.
\label{eq3.10}
\end{eqnarray}

\begin{eqnarray}
B\left\{-\left(1-\beta m^2\right) \sqrt {g^{\theta \theta}}\partial _ {\theta} I
+ \beta \sqrt {g^{\theta \theta}}\partial _ {\theta}I\left[g^{rr}(\partial _r I)^2 + g^{\theta \theta}(\partial _ {\theta}I)^2
+ g^{zz}(\partial _ {z}I)^2 \right]\right.\nonumber\\
\left.-i\left(1-\beta m^2\right) \sqrt {g^{zz}}\partial _ {z}I+ i\beta \sqrt {g^{zz}}\partial _ {z}I\left[g^{rr}
(\partial _r I)^2 + g^{\theta \theta}(\partial _ {\theta}I)^2 + g^{zz}(\partial _ {z}I)^2\right]\right\} = 0.
\label{eq3.11}
\end{eqnarray}

\noindent Obviously, it is difficult to get the solution of the action $I$ from the above equations. However, the action can be separated by the property of the black string. Considering the Killing vectors of the spacetime, the author separated the action as $I = -\omega t + W(r) + l\theta +Jz$ \cite{AS}, where $\omega$ is the energy of the emitted fermion. From the above four equations, we carry out separation of variables as

\begin{eqnarray}
I = -\omega t + W(r) + \Theta(\theta,z).
\label{eq3.12}
\end{eqnarray}

\noindent We first observe Eqs. (\ref{eq3.10}) and (\ref{eq3.11}) and find that they are irrelevant to $A$ and $B$ and can be reduced to the same equation. Inserting Eq. (\ref{eq3.12}) into Eqs. (\ref{eq3.10}) and (\ref{eq3.11}) yields

\begin{eqnarray}
\left(\sqrt{ g^{\theta \theta}}\partial _ {\theta}\Theta + i\sqrt {g^{zz}}\partial _ {z}\Theta\right)\left[1 - \beta m^2 -\beta g^{rr}
(\partial _r W)^2 - \beta g^{\theta \theta}(\partial _ {\theta}\Theta)^2 - \beta  g^{zz}(\partial _ {z}\Theta)^2\right] = 0.
\label{eq3.13}
\end{eqnarray}

\noindent In the above equation, the summation of factors in the square brackets can not be zero. Therefore, it should be

\begin{eqnarray}
\sqrt{ g^{\theta \theta}}\partial _ {\theta}\Theta + i\sqrt {g^{zz}}\partial _ {z}\Theta = 0,
\label{eq3.14}
\end{eqnarray}

\noindent which yields a complex function solution (other than the trivial constant solution) of $\Theta$. However, this solution has no contribution to the tunnelling rate. Therefore, we will not consider its contribution in the calculation. Another important relation predicted by Eq. (\ref{eq3.14}) is $g^{\theta \theta}(\partial _ {\theta}\Theta )^2 + g^{zz}(\partial _ {z}\Theta )^2 = 0$. Now we focus our attention on the first two equations. Inserting Eq. (\ref{eq3.12}) into Eqs. (\ref{eq3.8}) and (\ref{eq3.9}) and canceling $A$ and $B$ yield

\begin{eqnarray}
 A_6\left( {\partial _r W} \right)^6 + A_4\left({\partial _r W} \right)^4 + A_2\left( {\partial _r W} \right)^2 + A_0 =0,
\label{eq3.15}
\end{eqnarray}

\noindent where

\begin{eqnarray}
A_6 &=& \beta^{2}g^{3}f,\nonumber\\
A_4 &=& \beta g^{2}f\left(m^{2}\beta-2\right)-\beta ^2 g^2 e^2A_t^2,\nonumber\\
A_2 &=& gf\left(1-\beta m^{2}\right)\left(1+\beta m^{2}\right)-2\beta g eA_t [ \omega - eA_t(1-\beta m^2)],\nonumber\\
A_0 &=& -m^{2}f\left(1-\beta m^{2}\right)^{2}-\left[\omega -eA_t\left(1-\beta m^{2}\right)\right]^2.
\label{eq3.16}
\end{eqnarray}

\noindent Neglect the higher order terms of $\beta$ and solve Eq. (\ref{eq3.15}) at the event horizon. Thus the imaginary part of the radial action is

\begin{eqnarray}
ImW_{\pm} &=& \pm\int \frac{dr}{\sqrt{gf}} \sqrt{\left[ \omega - eA_t (1-\beta m^2)\right]^2+m^{2}f}\left(1+\beta m^{2} + \beta\frac{\tilde\omega_0^{2}- eA_t\tilde \omega_0}{f}\right)\nonumber\\
&=& \pm \pi\frac{\omega -eA_{t+}}{f' }\left(1+ \beta \xi \right),
\label{eq3.17}
\end{eqnarray}

\noindent where $+ (-)$ denote the outgoing (ingoing) solutions,  $f'=2\alpha^2r_+ + \frac{4M}{\alpha r_+^2} - \frac{8Q^2}{\alpha^2 r_+^3}$, $\xi = \frac{3}{2} m^2 + \frac{2m^2+1}{2\omega_0} +\frac{2eA_{t+}}{f'r_+} - \frac{2}{3}\frac{\omega_0}{f'r_+} $, $\tilde\omega_0 = \omega-eA_t$, $\omega_0 = \omega-eA_{t+}$, $A_{t+} = \frac{2Q}{\alpha r_+}$ is the electromagnetic potential at the event horizon. Using the relation of the roots of $f= 0$, it is easily proved that $\xi>0 $. The tunnelling rate of the charged fermion at the event horizon is

\begin{eqnarray}
\Gamma &\propto & exp[-Im\oint p_r dr] =exp\left[-Im\left(\int p_r^{out}dr-\int p_r^{in}dr\right)\right]\nonumber\\
&=& exp\left[\mp 2Im\int p_r^{out,in}dr\right] .
\label{eq3.18}
\end{eqnarray}

\noindent Here $p_r = \partial_r W$, and $out (in)$ correspond to $+ (-)$. Thus the tunnelling rate is gotten as

\begin{eqnarray}
\Gamma \propto \exp \left[-2\pi\frac{\omega -eA_{t+}}{f' }\left(1+ \beta \xi \right)\right].
\label{eq3.18-1}
\end{eqnarray}

\noindent However, the temporal contribution to the tunneling amplitude was missed in the above calculation \cite{APGS,ETA1,ETA2,ETA3}. We use Kruskal coordinates $(T, R)$ to find this temporal contribution. The region exterior is described by

\begin{eqnarray}
T &=& e^{\kappa r_*}sinh(\kappa t),\nonumber\\
R &=& e^{\kappa r_*}cosh(\kappa t),
\label{eq3.18-2}
\end{eqnarray}

\noindent  where $r_*= r+ \frac{1}{2\kappa}ln\frac{r-r_+}{r_+}$ is the tortoise coordinate, and $\kappa$ is the surface gravity. The interior region is given by

\begin{eqnarray}
T &=& e^{\kappa r_*}cosh(\kappa t),\nonumber\\
R &=& e^{\kappa r_*}sinh(\kappa t).
\label{eq3.18-3}
\end{eqnarray}

\noindent To find the temporal contribution, we connect these two patches across the horizon. Rotate the time $t$ as $t\rightarrow t- \frac{\pi}{2}i\kappa $. As pointed in \cite{APGS,APGS1,APGS2}, this ¡°rotation¡± would lead to an additional imaginary contribution coming from the temporal part, namely, $Im[(\omega - e A_{t+})\Delta t^{out,in}]=\frac{1}{2}\pi (\omega - e A_{t+})\kappa $. So the total temporal contribution is $Im[(\omega - e A_{t+})\Delta t]=\pi (\omega - e A_{t+})\kappa $.  Therefore, the tunnelling rate with the consideration of the temporal contribution is

\begin{eqnarray}
\Gamma &\propto & exp\left[-\frac{1}{\hbar}\left(Im ((\omega - e A_{t+})\Delta t)+Im\oint p_r dr\right)\right] \nonumber\\
&=& \exp \left[-4\pi\frac{\omega -eA_{t+}}{f' }\left(1+ \frac{1}{2}\beta \xi \right)\right].
\label{eq3.18-4}
\end{eqnarray}

\noindent This is the Boltzmann factor with the Hawking temperature at the event horizon taking

\begin{eqnarray}
T = \frac{f'}{4\pi \left(1+ \beta \xi \right)} = T_0 \left(1- \frac{1}{2}\beta \xi \right),
\label{eq3.19}
\end{eqnarray}

\noindent where $T_0  = \frac{1}{2\pi}\left(\alpha^2r_+ + \frac{2M}{\alpha r_+^2}-\frac{4Q^2}{\alpha^2 r_+^3}\right)$ is the standard Hawking temperature of the black string.

It is shown that the corrected temperature appears and is lower than the standard one. The correction is not only determined by the mass and charge of the black string, but also affected by the quantum number (mass, charge and energy) of the emitted fermion. Quantum gravity correction slows down the increase of the Hawking temperature caused by the evaporation. Finally, the black string is in a balance state. At this state, the evaporation stops and the remnant is produced.

It is of interest to discuss the corrected area entropy. The entropy can be derived by the first law of thermodynamics with the corrected temperature (\ref{eq3.19}). However, the expression is complicated, so we don't write it here. The corrected temperature were also gotten in\cite{BM07,BM08,BM09,BM10,BM11,BM12}. When $\beta = 0$, the standard Hawking temperature is recovered \cite{AS,CZ}.

\section{Fermions' tunnelling from a rotating black string}

In this section, we investigate uncharged fermions' tunnelling from the event horizon of a rotating black string. Therefore, effects of the electromagnetic field in the generalized Dirac equation are not taken into account here. The rotating black string solution in a spacetime asymptotically anti de Sitter in the radial direction was derived by Lemos \cite{LZ}. The solution is

\begin{eqnarray}
ds^2 &=& -\left(\alpha^2 r^2 - \frac{4M\left(1- \frac{a^2 \alpha^2}{2}\right)}{\alpha r}\right)dt^2 +\left(\alpha^2 r^2 - \frac{4M\left(1- \frac{3}{2}a^2 \alpha^2\right)}{\alpha r}\right)^{-1}dr^2\nonumber\\
&& - \frac{8Ma \sqrt{1- \frac{a^2 \alpha^2}{2}}}{\alpha r}dtd\varphi + \left( r^2 - \frac{4Ma^2}{\alpha r}\right)d\varphi^2+ \alpha^2 r^2 dz^2,
\label{eq4.1}
\end{eqnarray}

\noindent where $\alpha^2 = -\frac{\Lambda}{3} $, $\Lambda $ is the negative cosmological constant, $a$ is the angular momentum per unit mass. It is defined that  $a^2 \alpha^2 = 1- \frac{\epsilon}{M}$ and $\epsilon=\sqrt{M^2-\frac{8J^2\alpha^2}{9}}$. $M$ and $J$ are the mass and angular momentum line densities of the spacetime, respectively. The relation between $J$ and $a$ is given by $J= \frac{3}{2}Ma \sqrt{1- \frac{a^2 \alpha^2}{2}}$. For convenience of the investigation, the metric (\ref{eq4.1}) is rewritten as

\begin{eqnarray}
ds^2 &=& -\Delta \left(\gamma dt - \frac{\delta}{\alpha ^2}d\varphi\right)^2 + r^2\left(\gamma d\varphi - \delta dt\right)^2 +\frac{dr^2}{\Delta} + \alpha^2 r^2 dz^2,
\label{eq4.2}
\end{eqnarray}

\noindent where

\begin{eqnarray}
\Delta &=& \alpha^2 r^2 - \frac{b}{\alpha r},\quad b= 4M\left(1- \frac{3a^2 \alpha^2}{2}\right),\nonumber\\
\gamma &=& \sqrt{\frac{2-a^2\alpha^2}{2-3a^2\alpha^2}}, \quad \delta = \frac{a\alpha^2}{\sqrt{1-\frac{3}{2}a^2\alpha^2}}.
\label{eq4.3}
\end{eqnarray}

\noindent The event horizon is located at $r_+ = \alpha^{-1} b^{\frac{1}{3}} $ which is given for $\Delta = 0 $. To describe the fermion's tunnelling from the event horizon, one can directly construct a tetrad and gamma matrices from the metric (\ref{eq4.2}). For simplicity to construct the tetrad and gamma matrices, we perform the dragging coordinate transformation

\begin{eqnarray}
\varphi = \phi + \Omega t, \hspace{5mm}
\Omega = \frac{-\Delta \gamma  \delta\alpha^2 +r^2 \gamma \delta\alpha^4}{-\Delta \delta ^2 +r^2 \gamma^2 \alpha^4},
\label{eq4.4}
\end{eqnarray}

\noindent on the metric (\ref{eq4.2}) and get

\begin{eqnarray}
ds^2 &=& -F(r)dt^2 + \frac{1}{G(r)}dr^2+ g_{\phi\phi}d\phi^2 + g_{zz} dz^2\nonumber\\
&=& -\frac{\Delta r^2\left(\alpha^2\gamma^2-\delta^2\right)^2}{-\Delta \delta^2+\alpha^4r^2\gamma^2}dt^2+\frac{1}{\Delta}dr^2+\left(-\frac{\Delta\delta^4}{\alpha^4}+r^2\gamma^2\right)d\phi^2 +\alpha^2 r^2 dz^2.
\label{eq4.5}
\end{eqnarray}

\noindent Here we still only investigate the state with spin up. Assume that the wave function of the fermion with spin up state shares the similar expression as Eq . (\ref{eq3.5}), namely,

\begin{eqnarray}
\Psi=\left(\begin{array}{c}
A\\
0\\
B\\
0
\end{array}\right)\exp\left(\frac{i}{\hbar}I\left(t,r,\phi , z \right)\right).
\label{eq4.6}
\end{eqnarray}

\noindent The tetrad is easily constructed as

\begin{eqnarray}
e_\mu\,^a = \rm{diag}\left(\sqrt F, 1/\sqrt G, \sqrt {g_{\phi\phi}},\sqrt {g_{zz}} \right).
\label{eq4.7}
\end{eqnarray}

\noindent Now gamma matrices take on the form as

\begin{eqnarray}
\gamma^{t}=\frac{1}{\sqrt{F\left(r\right)}}\left(\begin{array}{cc}
i & 0\\
0 & -i
\end{array}\right), &  & \gamma^{\phi}=\sqrt{g^{\phi\phi}}\left(\begin{array}{cc}
0 & \sigma^{1}\\
\sigma^{1} & 0
\end{array}\right),\nonumber \\
\gamma^{r}=\sqrt{G\left(r\right)}\left(\begin{array}{cc}
0 & \sigma^{3}\\
\sigma^{3} & 0
\end{array}\right), &  & \gamma^{z}=\sqrt{g^{zz}}\left(\begin{array}{cc}
0 & \sigma^{2}\\
\sigma^{2} & 0
\end{array}\right).
\label{eq4.8}
\end{eqnarray}

\noindent In the above equations, $g^{\phi\phi} = \frac{\alpha^4}{-\Delta\delta^4+\alpha^4r^2\gamma^2}$,  $g^{zz} = \frac{1}{\alpha^2 r^2}$. Inserting the wave function and the gamma matrices into the generalized Dirac equation and adopting the same process as the above section, we get

\begin{eqnarray}
-iA\frac{1}{\sqrt{F}}\partial_{t}I-B\left(1-\beta m^{2}\right)\sqrt{G}\partial_{r}I-Am\beta\left[g^{rr}\left(\partial_{r}I\right)^{2}+
g^{\phi\phi}\left(\partial_{\phi}I\right)^{2}+g^{zz}\left(\partial_{z}I\right)^{2}\right]\nonumber\\
+B\beta\sqrt{G}\partial_{r}I\left[g^{rr}\left(\partial_{r}I\right)^{2}+ g^{\phi\phi}\left(\partial_{\phi}I\right)^{2} + g^{zz}\left(\partial_{z}I\right)^{2}\right]+Am\left(1-\beta m^{2}\right) = 0,
\label{eq4.9}
\end{eqnarray}

\begin{eqnarray}
iB\frac{1}{\sqrt{F}}\partial_{t}I-A\left(1-\beta m^{2}\right)\sqrt{G}\partial_{r}I-Bm\beta\left[g^{rr}\left(\partial_{r}I\right)^{2}+
g^{\phi\phi}\left(\partial_{\phi}I\right)^{2}+g^{zz}\left(\partial_{z}I\right)^{2}\right]\nonumber\\
+A\beta\sqrt{G}\partial_{r}I\left[g^{rr}\left(\partial_{r}I\right)^{2}+g^{\phi\phi}\left(\partial_{\phi}I\right)^{2} + g^{zz}\left(\partial_{z}I\right)^{2}\right]+Bm\left(1-\beta m^{2}\right) = 0,
\label{eq4.10}
\end{eqnarray}

\begin{eqnarray}
A\left\{-\left(1-\beta m^{2}\right)\sqrt{g^{\phi\phi}} \partial_{\phi}I+\beta\sqrt{g^{\phi\phi}}\partial _{\phi}I\left[g^{rr}\left(\partial_{r}I\right)^{2}+
g^{\phi\phi}\left(\partial_{\phi}I\right)^{2}+g^{zz}\left(\partial_{z}I\right)^{2}\right]\right.\nonumber\\
\left.-i\left(1-\beta m^{2}\right)\sqrt{g^{zz}}\partial_{z}I+i\beta\sqrt{g^{zz}}\partial _{z}I\left[g^{rr}\left(\partial_{r}I\right)^{2}+
g^{\phi\phi}\left(\partial_{\phi}I\right)^{2}+g^{zz}\left(\partial_{z}I\right)^{2}\right]\right\} = 0,
\label{eq4.11}
\end{eqnarray}

\begin{eqnarray}
B\left\{-\left(1-\beta m^{2}\right)\sqrt{g^{\phi\phi}} \partial_{\phi}I+\beta\sqrt{g^{\phi\phi}}\partial _{\phi}I\left[g^{rr}\left(\partial_{r}I\right)^{2}+
g^{\phi\phi}\left(\partial_{\phi}I\right)^{2}+g^{zz}\left(\partial_{z}I\right)^{2}\right]\right.\nonumber\\
\left.-i\left(1-\beta m^{2}\right)\sqrt{g^{zz}}\partial_{z}I+i\beta\sqrt{g^{zz}}\partial _{z}I\left[g^{rr}\left(\partial_{r}I\right)^{2}+
g^{\phi\phi}\left(\partial_{\phi}I\right)^{2}\right]+g^{zz}\left(\partial_{z}I\right)^{2}\right\} = 0.
\label{eq4.12}
\end{eqnarray}

\noindent It is also difficult to solve the action $I$ from the above equations. We first observe the last two equations. They can be reduced into the same equation and yields $\sqrt{g^{\phi\phi}} \partial_{\phi}I + i\sqrt{g^{zz}}\partial_{z}I=0$. This implies

\begin{eqnarray}
g^{\phi\phi}\left(\partial_{\phi}I\right)^{2}+ g^{zz}\left(\partial_{z}I\right)^{2} =0.
\label{eq4.13}
\end{eqnarray}

\noindent Now our interest is the first two equations which determine the Hawking temperature of the black string. Considering the properties of the metrics (\ref{eq4.1}) and (\ref{eq4.5}), we carry out separation of variables as

\begin{eqnarray}
I=-\left(\omega -j \Omega \right)t+ W\left(r, z \right) +j \phi,
\label{eq4.14}
\end{eqnarray}

\noindent where $\omega$ and $j$ are the energy and angular momentum of the emitted fermion, respectively. Inserting Eq. (\ref{eq4.14}) into Eqs. (\ref{eq4.9}) and (\ref{eq4.10}) and canceling $A$ and $B$ yield

\begin{eqnarray}
B_6\left( {\partial _r W} \right)^6 + B_4\left({\partial _r W} \right)^4 +B_2\left( {\partial _r W} \right)^2 + B_0 = 0,
\label{eq4.15}
\end{eqnarray}

\noindent where

\begin{eqnarray}
B_6 & = & \beta^{2}G^{3}F,\nonumber \\
B_4 & = & \beta G^{2}F\left(m^{2}\beta -2\right),\nonumber \\
B_2 & = & GF\left[\left(1-\beta m^{2}\right)^{2}+2\beta m^{2}\left(1-m^{2}\beta\right)\right],\nonumber \\
B_0 & = & -m^{2}\left(1-\beta m^{2}\right)^{2}F- \left(\omega - j\Omega\right)^{2}.
\label{eq4.16}
\end{eqnarray}

\noindent Neglecting the higher order terms of $\beta$ and solving Eq. (\ref{eq4.15}) at the event horizon, we get the solution of $W$. Thus the imaginary part of $W$ is

\begin{eqnarray}
Im W_{\pm} & = &\pm\int dr\sqrt{\frac{m^{2}F+\left(\omega - j\Omega\right)^{2}}{GF}} \left(1+\beta m^{2}+\beta\frac{\left(\omega - j\Omega\right)^{2}}{F}\right)\nonumber \\
 & = & \pm \pi \frac{\left(\omega - j \Omega_+\right)\alpha^4 r_+^2\gamma}{\left(2\alpha^4r_+^3 +b\right)\left(\alpha^2\gamma^2-\delta^2\right)}\left( 1+\beta \chi\right),
\label{eq4.17}
\end{eqnarray}

\noindent where $+ (-)$ are the outgoing (ingoing) solutions, $\chi= \frac{3}{2} m^2+\frac{3j\left(\omega -j\Omega_+\right)\delta} {\left(\alpha^2 \gamma^2-\delta^2\right) \gamma r_+^2}-\frac{3}{2} \frac{\left(\omega - j\Omega_+\right)^2\delta^2}{\left(\alpha^2\gamma^2-\delta^2\right) r_+^2}$, $\Omega_+ = \delta/\gamma$ is the angular velocity at the event horizon. It is not difficult to prove that $\chi>0$.

To find the temporal contribution, we use the Kruskal coordinates $(T,R)$. The region exterior to the string $(r>r_+)$ is described by

\begin{eqnarray}
T &=& e^{\kappa_+ r_*}sinh(\kappa_+ t),\nonumber\\
R &=& e^{\kappa_+ r_*}cosh(\kappa_+ t),
\label{eq4.17-1}
\end{eqnarray}

\noindent where $r_*= r+ \frac{1}{2\kappa_+}ln\frac{r-r_+}{r_+} $, and $\kappa_+ $ denote the surface gravity. The interior region is

\begin{eqnarray}
T &=& e^{\kappa_+r_*}cosh(\kappa_+t),\nonumber\\
R &=& e^{\kappa_+r_*}sinh(\kappa_+t).
\label{eq4.17-2}
\end{eqnarray}

\noindent Adopt the same process as the above section, we get the total temporal contribution is $Im[(\omega - j \Omega_+)\Delta t]=\pi (\omega - j \Omega_+)\kappa_+$.  Therefore, the tunnelling rate is

\begin{eqnarray}
\Gamma &\propto & exp\left[-\frac{1}{\hbar}\left(Im ((\omega - j \Omega_+)\Delta t)+Im\oint p_r dr\right)\right] \nonumber\\
&=& \exp \left[-\frac{4\pi \left(\omega - j \Omega_+\right)\alpha^4 r_+^2\gamma}{\left(2\alpha^4r_+^3 +b\right)\left(\alpha^2\gamma^2-\delta^2\right)}\left( 1 +\frac{1}{2} \beta \chi\right)\right].
\label{eq4.18}
\end{eqnarray}

\noindent Eq. (\ref{eq4.18}) is the Boltzmann factor of the Hawking temperature at the event horizon taking

\begin{eqnarray}
T &=& \frac{\left(2\alpha^4r_+^3 +b\right)\left(\alpha^2\gamma^2-\delta^2\right)}{4\pi\alpha^4 r_+^2\gamma\left( 1 +\frac{1}{2}\beta \chi\right)} =  T_0 \left(1 - \frac{1}{2}\beta \chi \right),
\label{eq4.19}
\end{eqnarray}

\noindent where $T_0 = \frac{\left(2\alpha^4r_+^3 +b\right)\left(\alpha^2\gamma^2-\delta^2\right)}{4\pi\alpha^4r_+^2\gamma}$ is the standard Hawking temperature. Obviously, the corrected Hawking temperature is lower than the standard one. The correction is related not only to the mass and angular momentum of the black string but also to the quantum number (mass, angular momentum and energy) of the emitted fermion. Due to $\chi>0$, there is a balance point. At this point, the evaporation stops and the remnant is left.

\section{Conclusion}

In this paper, taking into account the influence of quantum gravity, we modified the Dirac equation in curved spacetime by the modified fundamental commutation relations put forward in \cite{KMM}. Then the tunnelling radiation of fermions from the event horizons of the charged and rotating black strings was investigated. The corrected Hawking temperatures were gotten. In the charged spacetime, the correction is related not only to the mass and charge of the black string but also to the quantum number (mass, charge and energy) of the emitted fermion. In the rotating spacetime, the quantum number (mass, angular momentum and energy) of the emitted fermion and the mass and angular momentum of the black string affect the Hawking temperature at the same time. Due to the quantum gravity corrections, the evaporation of the black strings slows down. Finally, the evaporation stops and the remnants are left. The remnants in the final state were also discussed in \cite{ACS,SGC,BG,LX}. It is of interest to discuss remnants. A review of this topic can be found in \cite{N}.

When $\beta = 0$, the standard Hawking temperatures are recovered.

\vspace*{2.0ex}
\textbf{Acknowledgments}

This work is supported by the National Natural Science Foundation of China with Grant No. 11205125.


\bigskip

\begin{thebibliography}{99}

\small

\bibitem{KW}
P. Kraus and F. Wilczek, \emph{Self-interaction correction to  black hole radiance}, \emph{Nucl. Phys.} \textbf{B 433} (1995) 403 [arXiv:9408003].

\bibitem{APGS}
V. Akhmedova, T. Pilling, A. de Gill and D. Singleton, \emph{Temporal contribution to gravitational WKB-like calculations}, \emph{Phys. Lett.} \textbf{B 666} (2008) 269.

\bibitem{APGS1}
E.T. Akhmedov, T. Pilling and D. Singleton, \emph{Subtleties in the quasi-classical calculation of Hawking radiation}, \emph{Int. J. Mod. Phys.} \textbf{D 17} 322 (2008) 2453.

\bibitem{APGS2}
V. Akhmedova, T. Pilling, A. de Gill and D. Singleton, \emph{Comments on anomaly versus WKB/tunneling methods for calculating Unruh radiation}, \emph{Phys. Lett.} \textbf{B 673} (2009) 227.

\bibitem{ETA1}
E.T. Akhmedov, V. Akhmedova, T. Pilling and D. Singleton, \emph{Thermal radiation of various gravitational backgrounds}, \emph{Int. J. Mod. Phys.} \textbf{A 22} (2007) 1705.

\bibitem{ETA2}
B.D. Chowdhury, \emph{Problems with tunneling of thin shells from black holes}, \emph{Pramana} \textbf{70} (2008) 593.

\bibitem{ETA3}
E.T. Akhmedov, V. Akhmedova and D. Singleton, \emph{Hawking temperature in the tunneling picture}, \emph{Phys. Lett.} \textbf{B 642} (2006) 124.

\bibitem{PW}
M.K. Parikh and F. Wilczek, \emph{Hawking radiation as tunneling}, \emph{Phys. Rev. Lett.} \textbf{85} (2000) 5042 [arXiv:9907001].

\bibitem{SP}
K. Srinivasan and T. Padmanabhan, \emph{Particle production and complex path analysis}, \emph{Phys. Rev.} \textbf{D 60} (1999) 024007.

\bibitem{SPS}
S. Shankaranarayanan, T. Padmanabhan and K. Srinivasan, \emph{Hawking radiation in different coordinate settings: Complex paths approach}, \emph{Class. Quant. Grav.} \textbf{19} (2002) 2671.

\bibitem{ANVZ}
M. Agheben, M. Nadalini, L. Vanzo and S. Zerbini, \emph{Hawking radiation as tunneling for extremal and rotating black holes}, \emph{JHEP} \textbf{0505} (2005) 014 [arXiv:0503081].

\bibitem{KM1}
R. Kerner and R.B. Mann, \emph{Tunnelling, Temperature and Taub-NUT Black Holes}, \emph{Phys. Rev.} \textbf{D 73} (2006) 104010 [arXiv:0603019].

\bibitem{ZZ}
J.Y. Zhang and Z. Zhao, \emph{Hawking radiation of charged particles via tunneling from the Reissner-Nordstrom black hole}, \emph{JHEP} \textbf{10}
(2005) 055.

\bibitem{JWC}
Q.Q. Jiang and S.Q. Wu, \emph{Hawking radiation of charged particles as tunneling from Reissner-Nordstrom-de Sitter black holes with a global monopole}, \emph{Phys. Lett.} \textbf{B 635} (2006) 151 [arXiv:0511123]

\bibitem{KM2}
R. Kerner and R.B. Mann, \emph{Fermions tunnelling from black holes}, \emph{Class. Quant. Grav.} \textbf{25} (2008) 095014 [arXiv:0710.0612].

\bibitem{KM3}
R. Kerner and R.B. Mann, \emph{Charged Fermions Tunnelling from Kerr-Newman Black Holes}, \emph{Phys. Lett.} \textbf{B 665} (2008) 277 [arXiv:0803.2246].

\bibitem{AS}
J. Ahmed and K. Saifullah, \emph{Hawking temperature of rotating charged black strings from tunneling}, \emph{JCAP} \textbf{11} (2011) 023 [arXiv:1109.1710].

\bibitem{AS1}
J. Ahmed and K. Saifullah, \emph{Hawking radiation of Dirac particles from black strings}, \emph{JCAP} \textbf{08} (2011) 011 [arXiv:1108.2677].

\bibitem{LR}
R. Li, J.R. Ren, \emph{Dirac particles tunneling from BTZ black hole}, \emph{Phys. Lett.} \textbf{B 661} (2008) 370 [arXiv:0802.3954].

\bibitem{CJZ}
D.Y. Chen, Q.Q. Jiang and X.T. Zu, \emph{Hawking radiation of Dirac particles via tunnelling from rotating black holes in de Sitter spaces}, \emph{Phys. Lett.} \textbf{B 665} (2008) 106 [arXiv: 0804.0131].

\bibitem{QQJ}
Q.Q. Jiang, \emph{Dirac particles' tunnelling from black rings}, \emph{Phys. Rev.} \textbf{D 78} (2008) 044009 [arXiv:0807.1358].

\bibitem{HCNVZ}
R. Di Criscienzo and L. Vanzo, \emph{Fermion tunneling from dynamical horizons}, \emph{Eur. Phys. Lett.} \textbf{82} (2008) 60001 [arXiv:0803.0435].

\bibitem{HCNVZ1}
S.A. Hayward, R. Di Criscienzo, M. Nadalini, L. Vanzo and S. Zerbini, \emph{Local Hawking temperature for dynamical black holes}, [arXiv:0806.0014].

\bibitem{BRM}
B.R. Majhi, \emph{Fermion tunneling beyond semiclassical approximation}, \emph{Phys. Rev.} \textbf{D 79}  (2009) 044005 [arXiv:0809.1508].

\bibitem{RW}
S.P. Robinson and F. Wilczek, \emph{Relationship between Hawking radiation and gravitational anomalies}, \emph{Phys. Rev. Lett.} \textbf{95} (2005) 011303.

\bibitem{RW1}
S. Iso, H. Umetsu and F. Wilczek, \emph{Hawking radiation from charged black holes via gauge and gravitational anomalies}, \emph{Phys. Rev. Lett.} \textbf{96} (2006) 151302.

\bibitem{BM01}
R. Banerjee and B. R. Majhi, \emph{Connecting anomaly and tunneling methods for Hawking effect through chirality}, \emph{Phys. Rev.} \textbf{D 79} (2009) 064024.

\bibitem{PKTC}
P.K. Townsend, \emph{Small-scale structure of spacetime as  the origin of the gravitational constant}, \emph{Phys. Rev.} \textbf{D 15}
(1977) 2795.

\bibitem{PKTC1}
D. Amati, M. Ciafaloni and G. Veneziano, \emph{Can spacetime be probed below the string size?} \emph{Phys. Lett.} \textbf{B 216} (1989) 41.

\bibitem{PKTC2}
K. Konishi, G. Paffuti and P. Provero, \emph{Minimum  physical length and the generalized uncertainty principle in string theory}, \emph{Phys. Lett.} \textbf{B 234} (1990) 276.

\bibitem{PKTC3}
L.J. Garay, \emph{Quantum gravity and minimum length}, \emph{Int. J. Mod. Phys.} \textbf{A 10} (1995) 145 [arXiv:9403008].

\bibitem{PKTC4}
G. Amelino-Camelia, \emph{Relativity in space-times with short-distance structure governed by an observer-independent (Planckian) length scale}, \emph{Int. J. Mod. Phys.} \textbf{D 11} (2002) 35 [arXiv:0012051].

\bibitem{KMM}
A. Kempf, G. Mangano and R.B. Mann, \emph{Hilbert space representation of the minimal length uncertainty relation}, \emph{Phys. Rev.} \textbf{D 52} (1995) 1108 [arXiv:9412167].

\bibitem{CMOT}
L.N. Chang, D. Minic, N. Okamura and T. Takeuchi, \emph{The effect of the minimal length uncertainty relation on the density of states and the cosmological constant problem}, \emph{Phys. Rev.} \textbf{D 65} (2002) 125028 [arXiv:0201017].

\bibitem{BM06}
B.R. Majhi and E.C. Vagenas, \emph{Modified dispersion relation, photon's velocity, and Unruh effect}, \emph{Phys. Lett.} \textbf{B 725} (2013) 477.

\bibitem{BM1}
M.V. Battisti and G. Montani, \emph{The Big-Bang Singularity in the framework of a Generalized Uncertainty Principle}, \emph{Phys. Lett.} \textbf{B 656} (2007) 96 [arXiv:0703025].

\bibitem{CDAV}
W. Chemissany, S. Das, A.F. Ali and E.C. Vagenas, \emph{Effect of the Generalized Uncertainty Principle on post-inflation preheating}, \emph{JCAP} 1112 (2011) 017 [arXiv:1111.7288].

\bibitem{KSY}
W. Kim, E. J. Son and M. Yoon, \emph{Thermodynamics of a black hole based on a generalized uncertainty principle}, \emph{JHEP} \textbf{0801} (2008) 035.

\bibitem{XW}
L. Xiang and X.Q. Wen, \emph{Black hole thermodynamics with generalized uncertainty principle}, \emph{JHEP} \textbf{0910} (2009) 046 [arXiv:0901.0603[qr-qc]].

\bibitem{BJM}
A. Bina, S. Jalalzadeh and A. Moslehi, \emph{Quantum black hole in the generalized uncertainty principle framework}, \emph{Phys. Rev.} \textbf{D 81} (2010) 023528 [arXiv:1001.0861[qr-qc]].

\bibitem{NS}
K. Nozari and S. Saghafi, \emph{Natural cutoffs and quantum tunneling from black hole horizon}, \emph{JHEP} \textbf{11} (2012) 005 [arXiv:1206.5621].

\bibitem{CWY}
D.~Chen, H.~Wu and H.~Yang, \emph{Fermion's tunnelling with effects of quantum gravity}, \emph{Advances in High Energy Physics} \textbf{2013} (2013) 432412.

\bibitem{WG}
W. Greiner, Relativistic Quantum Mechanics: Wave Equation, Springer-Verlag, 2000.

\bibitem{NK1}
K. Nozari and M. Karami, \emph{Minimal length and generalized Dirac equation}, \emph{Mod. Phys. Lett.} \textbf{A 20} (2005) 3095 [arXiv:0507028].

\bibitem{HBHRSS}
S. Hossenfelder, M. Bleicher, S. Hofmann, et. al., \emph{Signatures in the planck regime}, \emph{Phys. Lett.} {\bf B 575} (2003) 85 [arXiv:0305262].

\bibitem{JPSL}
J.P.S. Lemos, \emph{Cylindrical black hole in general relativity}, \emph{Phys. Lett.} \textbf{B 353} (1995) 46 [arXiv:9404041].

\bibitem{CZ}
R.G. Cai and Y.Z. Zhang, \emph{Black plane solutions in four-dimensional spacetimes}, \emph{Phys. Rev.} \textbf{D 54} (1996) 4891 [arXiv:9609065].

\bibitem{BM07}
R. Banerjee and B.R. Majhi, \emph{Quantum tunneling and back reaction}, \emph{Phys. Lett.} \textbf{B 662} (2008) 62.

\bibitem{BM08}
R. Banerjee, B.R. Majhi, and S. Samanta, \emph{Noncommutative black hole thermodynamics}, \emph{Phys. Rev.} \textbf{D 77} (2008) 124035.

\bibitem{BM09}
R. Banerjee and B.R. Majhi, \emph{Quantum tunneling beyond semiclassical approximation}, \emph{JHEP} \textbf{0806} (2008) 095.

\bibitem{BM10}
R. Banerjee and B.R. Majhi, \emph{Quantum tunneling and trace anomaly}, \emph{Phys. Lett.} \textbf{B 674} (2009) 218.

\bibitem{BM11}
B.R. Majhi and S. Samanta, \emph{Hawking radiation due to photon and gravitino tunneling}, \emph{Annals Phys.} \textbf{325} (2010) 2410.

\bibitem{BM12}
R. Banerjee, C. Kiefer and B.R Majhi, \emph{Quantum gravitational correction to the Hawking temperature from the Lemaitre-Tolman-Bondi model}, \emph{Phys. Rev.} \textbf{D 82} (2010) 044013.

\bibitem{LZ}
J.P.S. Lemos and V.T. Zanchin, \emph{Rotating charged black string and three dimensional black holes}, \emph{Phys. Rev.} \textbf{D 54} (1996) 3840.

\bibitem{ACS}
R.J. Adler, P. Chen and D.I. Santiago, \emph{The generalized uncertainty principle and black hole remnants}, \emph{Gen. Rel. Grav.} \textbf{33} (2001) 2101 [arXiv:gr-qc/0106080].

\bibitem{SGC}
F. Scardigli, C. Gruber, P. Chen, \emph{Black hole remnants in the early universe}, \emph{Phys. Rev.} \textbf{D 83} (2011) 063507 [arXiv:1009.0882[qr-qc]].

\bibitem{BG}
R. Banerjee, S. Ghosh, \emph{Generalised uncertainty principle, remnant mass and singularity problem in black hole thermodynamics}, \emph{Phys. Lett.} \textbf{B 688} (2010) 224 [arXiv:1002.2302[gr-qc]].

\bibitem{LX}
L. Xiang, \emph{A note on the black hole remnant}, \emph{Phys. Lett.} \textbf{B 647} (2007) 207 [arXiv:gr-qc/0611028].

\bibitem{N}
P. Nicolini, \emph{Noncommutative black holes, the final appeal to quantum gravity: A review}, \emph{Int. J. Mod. Phys.} \textbf{A 24} (2009) 1229.




\end{thebibliography}
\end{document}